\documentclass[aps,prl,twocolumn,superscriptaddress,amsmath,amssymb]{revtex4-1}
\usepackage{url}
\usepackage{witharrows}
\usepackage{hyperref}
\usepackage{mathtools}
\usepackage{amsmath}
\usepackage{tensor}
\usepackage{palatino}
\usepackage{tikz}
\usepackage{multirow}
\usepackage{tabularx}
\usepackage{bm}
\usepackage{bbm}
\usepackage{pgfplots}
\usepackage{txfonts}
\usepackage{verbatim}
\usepackage{slashed}
\usepackage{subfig}
\usepackage{graphicx}
\usepackage{pstricks}
\usepackage{listings}
\usepackage{stackengine}
\usepgfplotslibrary{patchplots}
\pgfplotsset{compat=1.15}
\usetikzlibrary{backgrounds,shadows.blur,fit,decorations.pathreplacing,shapes.geometric,shapes.arrows,patterns, positioning, arrows, intersections, fadings, decorations.pathreplacing}
\usetikzlibrary{quotes,angles}
\usetikzlibrary{calc}
\usetikzlibrary{decorations.pathreplacing}
\usetikzlibrary{arrows.meta}
\tikzset{global scale/.style={
		scale=#1,
		every node/.append style={scale=#1}}}

\begin{document}

\title{Cosmological Constant, Inflaton, and Dark Matter all Naturally Originated from Poincar\'e Gauge Gravity}

\author{Hongchao Zhang}
\email{zhanghongchao852@live.com}
\affiliation{Institute for Theoretical Physics \& Cosmology, Zhejiang University of Technology, Hangzhou, 310023, China\\
United Center for Gravitational Wave Physics (UCGWP), Zhejiang University of Technology, Hangzhou, 310023, China}

\author{Tao Zhu}
\email{zhut05@zjut.edu.cn}
\affiliation{Institute for Theoretical Physics \& Cosmology, Zhejiang University of Technology, Hangzhou, 310023, China\\
United Center for Gravitational Wave Physics (UCGWP), Zhejiang University of Technology, Hangzhou, 310023, China}

\author{Anzhong Wang}
\email{anzhong$\_$wang@baylor.edu}
\affiliation{GCAP-CASPER, Physics Department, Baylor University, Waco, Texas 76798-7316, USA}

\date{\today}

\begin{abstract}
We propose a cosmological model in the framework of Poincar\'e gauge gravity, in which cosmological constant, inflaton, and dark matter candidate all naturally originate.
Cosmological constant originates in the process of breaking of the Poincar\'e symmetries down to the Lorentz symmetries.
We select a gauge Lagrangian without any additional matter fields, which can be regarded as a minimum extension of general relativity with two more massive modes from the Lorentz connection.
Numerical analysis shows that the scalar dominates a slow-rolling inflation and the pseudo-scalar behaves as a dark matter candidate.

\end{abstract}


\maketitle

Cosmic inflation, dark matter and late-time acceleration, are three main tensions between modern cosmology based on Einstein's general relativity (GR) and observations\cite{vazquez2018inflationary,bertone2018history,frieman2008dark,peebles2003cosmological}.
In recent decades, a series of models based on the extension of the standard model of particle physics and/or the modification of GR have been proposed to address these tensions.
Some of these models have been ruled out by observational data, while others are still undergoing further rigorous tests from both theory and observation\cite{yoo2012theoretical,vennin2015cosmic,young2017survey,saridakis2021modified}.
So far, the optimal models for inflation, dark matter and late-time acceleration in line with observations are slow-rolling inflation dominated by a single scalar field, cold dark matter particles and cosmological constant, respectively.
However, explaining these three phenomena within a unified theoretical framework that is compatible with both the standard model and gravitational theory remains a long-term and challenging task.
In the standard model, $U(1)\times SU(2)\times SU(3)$ gauge theories describe the generations of electromagnetic, weak, and strong interactions, respectively. Spontaneous symmetry breaking combined with the Higgs mechanism explains the source of the masses of gauge bosons.
From this perspective, gravity theories by gauging spacetime groups have more advantages in compatibility with the standard model than GR.
The Poincar\'e group is the maximum isometric group in Minkowski spacetime, and also the representation group of elementary particles. Localization of this group leads us to the Poincar\'e gauge theory of gravity (PGG)\cite{hehl1976general,ivanenko1983gauge,hehl1995metric,hehl2017gauge}. Due to the fact that the Poincar\'e group involves the translations between different points in spacetime (which leads to the Poincar\'e group being external), we cannot simply define gauge transformation as vertical automorphism along fibre, as in those gauge theories of internal groups, but rather introduce automorphisms between fibres in a nonlinear way\cite{hennig1981gravity,lord1988gauge,tresguerres2002unified,tiemblo2005gauge}.
Coincidentally, nonlinear representation is not only an appropriate approach to describe the gauge theories of spacetime groups, but also introduces the natural description of spontaneous symmetry breaking\cite{stelle1980spontaneously,ne1988gravity,tresguerres2000gravitational,kirsch2005higgs,leclerc2006higgs}. Then, by introducing the appropriate Higgs mechanisms, the story of gravity will naturally develop in the same direction as the standard model.

The goal of this letter is to introduce a cosmological scenario in which the cosmological constant, inflaton and (cold) dark matter candidate naturally originate in the framework of PGG.

\textit{Origination of cosmological constant from the breakdown of the Poincar\'e symmetries to the Lorentz symmetries.}---
From the perspective of gauge field theory, a Poincar\'e (P-) observer (base) at point $x$ can be expressed by a binary tuple $\left\{ \begin{array}{cc} e& v \end{array} \right\}_x$, with $e$ a Lorentz (L-) observer and $v$ a base for measuring ``internal coordinates''\cite{kirsch2005higgs}.
By introducing a five dimensional matrix representation\cite{hennig1981gravity,leclerc2006higgs}, a P-transformation on a P-observer can be performed in the following way
\begin{equation}
	\left\{ \begin{array}{cc}
		\tilde{e}& \tilde{v}\end{array} \right\}_x=\left\{ \begin{array}{cc}
		e& v\end{array} \right\}_x
	\left(\begin{array}{cc}
		\Lambda&\xi\\ 0&1\end{array}\right),
\end{equation}
where $\Lambda$ is an element of the L-group, and $\xi\in\mathbb{R}^4$ is an element of the translational group.
A P-field of matter (referred to the definition of a matter field in fibre bundle terminology---a section in the fibre bundle associated to principal fibre bundle) is an equivalence class with respect to P-observer given by
\begin{equation}
	\Psi=\left(\left\{ \begin{array}{cc}
		e& v \end{array} \right\},\left(\begin{array}{cc}
		\psi&y\\ 0&1 \end{array}\right)
	\right)/\sim,
\end{equation}
where $\psi$ is a L-field in a certain representation, and $y$ is the corresponding ``internal coordinates'' measured by $v$.
The Localization of the P-symmetries leads to the introduction of P-gauge field (connection) and the P-covariant derivative
\begin{equation}
	\bm{\lozenge}\equiv\left(\begin{array}{cc}
		\bm{\mathcal{D}}&\bm{B}\\ 0&0\end{array}\right),
\end{equation}
where $\bm{\mathcal{D}}\equiv\bm{d}+\bm{A}$ is the L-covariant derivative with respect to the L-connection $\bm{A}$. According to the transformation properties\cite{lord1988gauge,leclerc2006higgs}, $\bm{B}$ is a connection-like P-vector.
The action of $\bm{\lozenge}$ on a P-field, i.e. P-velocity of a P-field is
\begin{equation}\label{def:P-velocity}
	\bm{\lozenge}\left(\begin{array}{cc}
		\psi&y\\ 0&1\end{array}\right)=\left(\begin{array}{cc}
		\bm{\mathcal{D}}\psi&\bm{\theta}\\ 0&0 \end{array}\right),
\end{equation}
where $\bm{\theta}\equiv\bm{\mathcal{D}}y+\bm{B}$ is the so called canonical $1$-form, which is a L-vector according to the transformation properties.
The P-gauge strength can be obtained by twice actions of $\bm{\lozenge}$
\begin{equation}\label{def:gauge_strength}
	\bm{\lozenge}\wedge\bm{\lozenge}\equiv\left(\begin{array}{cc}
		\bm{R}&\bm{S}\\ 0&0 \end{array}\right),
\end{equation}
where $\bm{R}$ is curvature, and $\bm{S}$ is a gauge strength-like P-vector.
Forthermore, the action of (\ref{def:gauge_strength}) on a P-field read
\begin{equation}\label{def:gauge_torsion}
	\bm{\lozenge}\wedge\bm{\lozenge}\left(\begin{array}{cc}
		\psi&y\\ 0&1 \end{array}\right)=\left(\begin{array}{cc}
		\bm{R}\psi&\bm{T}\\ 0&0 \end{array}\right),
\end{equation}
where $\bm{T}\equiv\bm{R}y+\bm{S}$ is torsion---a L-vector.

In order to construct the Lagrangian of a P-field, including the kinetic energy term and the interaction with P-gauge field, a common approach is
\begin{equation}\label{def:invalid_matter}
	L_M=\bm{\lozenge}\left(\begin{array}{cc}
		\psi&y\\ 0&1\end{array}\right)\cdot\bm{\lozenge}\left(\begin{array}{cc}
		\psi&y\\ 0&1\end{array}\right).
\end{equation}
However, because P-algebra is not semi-simple, it's impossible to define such an inner product ``$\cdot$''\cite{leclerc2006higgs}, so the expression in (\ref{def:invalid_matter}) is invalid.
However, there exist a Killing-Cartan metric, i.e. the so called Minkowski metric $\eta$ in L-algebra, so that the inner product between L-vectors can be defined. It so happens that the two components $\bm{\mathcal{D}}\psi$ and $\bm{\theta}$ in the r.h.s. of P-velocity (\ref{def:P-velocity}) are quantities with L-representation. Therefore, the Lagrangian in (\ref{def:invalid_matter}) can be modified to
\begin{equation}\label{def:valid_matter}
	L_M=\eta(\bm{\mathcal{D}}\psi,\bm{\mathcal{D}}\psi)+\eta(\bm{\theta},\bm{\theta}).
\end{equation}
The first term in the r.h.s of (\ref{def:valid_matter}) is the kinetic energy term of L-field, including the interaction with L-gauge field. While the second term can be rewritten in\cite{stelle1980spontaneously,leclerc2006higgs} (release all hidden indices)
\begin{equation}\label{eq:eta4}
	\eta(\bm{\theta},\bm{\theta})=g^{\mu\nu}e_\mu{}^ae_\nu{}^b\eta_{ab}=4,
\end{equation}
with $e_\mu{}^a=\partial_\mu\lrcorner\bm{\theta}^a$ referred to the tetrad or vielbein field.
This means that in the process of breaking the P-symmetries down to the L-symmetries, a constant term naturally originates in the Lagrangian of matter.

On the other hand, the Lagrangian $4$-form of the P-gauge field would be
\begin{equation}\label{def:invalid_gauge}
	L_G=\left(\begin{array}{cc}
		\bm{R}&\bm{S}\\ 0&0\end{array}\right)\wedge^*\left(\begin{array}{cc}
		\bm{R}&\bm{S}\\ 0&0\end{array}\right).
\end{equation}
But due to the same reason as above, expression (\ref{def:invalid_gauge}) is invalid. By means of L-metric $\eta$, it can be modified in
\begin{equation}\label{def:valid_gauge}
	L_G=\bm{R}\wedge^*\bm{R}+\bm{T}\wedge^*\bm{T}	
\end{equation}
according to (\ref{def:gauge_torsion}), which is the general form of the Yang-Mills (YM) type gauge Lagrangian for ``Poincar\'e'' gauge gravity that we are familiar with. Obviously, both $\bm{R}$ and $\bm{T}$ are L-tensors, thus (\ref{def:valid_gauge}) is the expression after the breaking of the P-symmetries down to the L-symmetries.

To summarize this section, in the framework of PGG, the action including the matter field and the gauge field should take the following form
\begin{alignat}{2}
	s=s_M+s_G=&\int dx^4e(\mathcal{D}_\mu\psi\mathcal{D}^\mu\psi+\lambda)\nonumber\\
	&+\int dx^4e[\tfrac{1}{2\kappa}(R+T^2)+R^2],
\end{alignat}
with $e\equiv\det(e_\mu{}^a)$ and $\lambda$ a constant plugging (\ref{eq:eta4}) in and $\kappa\sim m_{Pl}^{-2}$ the coupling constant of gravity.
In the gauge action, in addition to the quadratic terms of the field strengthes, we also consider the linear curvature term, namely the Einstein-Hilbert (EH) term. For simplicity, we only consider the parity-conserving terms.


\textit{A minimum model.}---
Generally speaking, the quadratic terms in the gauge Lagrangian can be decomposed into several inequivalent and irreducible pieces\cite{baekler2011beyond}, and the system may contain ghosts and tachyons.
According to \cite{neville1980gravity,sezgin1980new,lin2019ghost,blagojevic2018general,karananas2015particle,lin2020power,zhang2019late,zhang2019inflation}, the most general ghost- and tachyon-free YM type Lagrangian of PGG contains up to eight kinds of \emph{possible} modes in terms of the $SO(3)$ spin-parity decomposition: two massless and six massive. However, except for one massless spin-$0^+$ mode and two massive spin-$0^\pm$ modes, none of them are of concern to us in the following cosmological context.
Fortunately, those modes can be suppressed by selecting appropriate combinations of Lagrangian parameters.
As a result, we selected a minimum ghost- and tachyon-free parity-conserving EH-YM type Lagrangian based on \cite{yo1999hamiltonian,yo2002hamiltonian} as follows
\begin{alignat}{2}\label{eq:Lagrangian}
	L_G=&b_0R+\tfrac{b_0}{3}T_{\mu\nu\rho}(T^{\mu\nu\rho}+T^{\rho\nu\mu}-g^{\mu\rho}T^{\nu})\nonumber\\
	&+\tfrac{2A_1}{3}T_{\mu}T^{\mu}+\tfrac{A_2}{12}T_{\mu\nu\rho}(2T^{\rho\nu\mu}-T^{\mu\nu\rho})\nonumber\\
	&+\tfrac{B_1}{9}(R_{\mu\nu}R^{\nu\mu}-\tfrac{1}{4}R_{\mu\nu\rho\sigma}R^{\rho\sigma\mu\nu})\nonumber\\
	&+\tfrac{B_2}{9}R_{\mu\nu\rho\sigma}(R^{\mu\rho\nu\sigma}-\tfrac{1}{4}R^{\mu\nu\rho\sigma}-\tfrac{1}{4}R^{\rho\sigma\mu\nu}),
\end{alignat}
where $b_0$, $A_1$, $A_2$ are parameters in the unit of quadratic Planck mass, i.e. $m_{Pl}^2$, and $B_1$, $B_2$ are dimensionless. All parameters are non-zero positive.
The combination of terms in (\ref{eq:Lagrangian}) not only guarantees that there are no ghosts and tachyons, but also no extra modes, such as spin-$1^{\pm}$, spin-$2^-$ and massive spin-$2^{+}$, up to the linear perturbed order. It should be emphasized that this Lagrangian still keeps a massless spin-$2^+$ mode with same propagator as the conversant gravitational waves in GR.
Therefore, the Lagrangian (\ref{eq:Lagrangian}) can be regarded as a minimum extension of GR within the framework of PGG from the perspective of ``particles''. See FIG. \ref{fig:spectrum} for a more intuitive display.
Studies of the similar Lagrangian and their cosmological applications in various scenarios, please see \cite{yo2007dynamic,shie2008torsion,baekler2011poincare}.
\tikzset{%
	brace/.style = { decorate, decoration={brace, amplitude=5pt} },
	mbrace/.style = { decorate, decoration={brace, amplitude=5pt, mirror} },
	label/.style = { black, midway, scale=0.6, align=center },
	toplabel/.style = { label, above=.6em, anchor=south },
	leftlabel/.style = { label,rotate=-90,left=.6em,anchor=north },   
	bottomlabel/.style = { label, below=.6em, anchor=north },
	force/.style = { rotate=-90,scale=0.5 },
	round/.style = { rounded corners=2mm },
	legend/.style = { right,scale=0.5 },
	nosep/.style = { inner sep=0pt },
	generation/.style = { anchor=base }
}
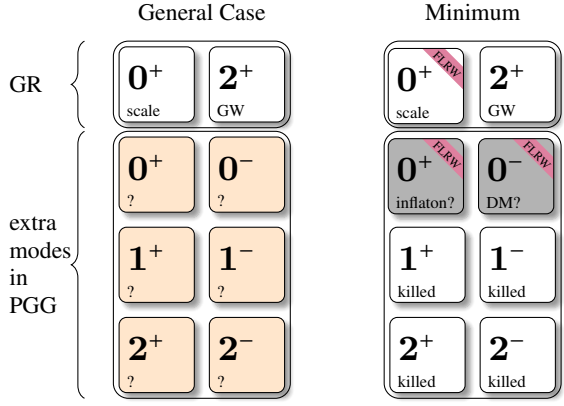
\begin{figure}[htbp]
	\centering
	\newcommand\particle[7][white]{%
		\begin{tikzpicture}[x=1cm, y=1cm]
			\path[fill=#1,blur shadow={shadow blur steps=5}] (0.1,0) -- (0.9,0)
			arc (90:0:1mm) -- (1.0,-0.9) arc (0:-90:1mm) -- (0.1,-1.0)
			arc (-90:-180:1mm) -- (0,-0.1) arc(180:90:1mm) -- cycle;
			\ifstrempty{#7}{}{\path[fill=purple!50!white]
				(0.45,0) --(0.75,0) -- (1.0,-0.25) -- (1.0,-0.55);}
			\ifstrempty{#6}{}{\path[fill=green!50!black!50] (0.7,0) -- (0.9,0)
				arc (90:0:1mm) -- (1.0,-0.3);}
			\ifstrempty{#5}{}{\path[fill=orange!50!white] (1.0,-0.7) -- (1.0,-0.9)
				arc (0:-90:1mm) -- (0.7,-1.0);}
			\draw[\ifstrempty{#2}{dashed}{black}] (0.1,0) -- (0.9,0)
			arc (90:0:1mm) -- (1.0,-0.9) arc (0:-90:1mm) -- (0.1,-1.0)
			arc (-90:-180:1mm) -- (0,-0.1) arc(180:90:1mm) -- cycle;
			\ifstrempty{#7}{}{\node at(0.78,-0.22) [rotate=-45,scale=0.51] {#7};}
			\ifstrempty{#6}{}{\node at(0.9,-0.1)  [nosep,scale=0.17] {#6};}
			\ifstrempty{#5}{}{\node at(0.9,-0.9)  [nosep,scale=0.2] {#5};}
			\ifstrempty{#4}{}{\node at(0.1,-0.1)  [nosep,anchor=west,scale=0.25]{#4};}
			\ifstrempty{#3}{}{\node at(0.1,-0.85) [nosep,anchor=west,scale=0.7] {#3};}
			\ifstrempty{#2}{}{\node at(0.1,-0.4)  [nosep,anchor=west,scale=1.5] {#2};}
		\end{tikzpicture}
	}
	
	\begin{tikzpicture}[x=1.2cm, y=1.2cm]
		\draw[round] (-0.48,0.48) rectangle (1.48,-0.48);
		\draw[round] (-0.48,-0.52) rectangle (1.48,-3.48);
		
		\node at(0, 0)   {\particle
			{$\bm{0}^+$}{scale}{}{}{}{}};
		\node at(0,-1)   {\particle[orange!20!white]
			{$\bm{0}^+$} {?}    {}{}{}{}};
		\node at(0,-2)   {\particle[orange!20!white]
			{$\bm{1}^+$}        {?}       {}{}{}{}};
		\node at(0,-3)   {\particle[orange!20!white]
			{$\bm{2}^+$}    {?}         {}{}{}{}};
		\node at(1, 0)   {\particle
			{$\bm{2}^+$}        {GW}   {}{}{}{}};
		\node at(1,-1)   {\particle[orange!20!white]
			{$\bm{0}^-$}        {?}  {}{}{}{}};
		\node at(1,-2)   {\particle[orange!20!white]
			{$\bm{1}^-$}      {?}        {}{}{}{}};
		\node at(1,-3)   {\particle[orange!20!white]
			{$\bm{2}^-$}  {?}    {}{}{}{}};
		
		\draw[round] (2.52,0.48) rectangle (4.48,-0.48);
		\draw[round] (2.52,-0.52) rectangle (4.48,-3.48);
		
		\node at(3, 0)   {\particle
			{$\bm{0}^+$}        {scale}       {}{}{}{FLRW}};
		\node at(3,-1)   {\particle[gray!60!white]
			{$\bm{0}^+$}        {inflaton?}    {}{}{}{FLRW}};
		\node at(3,-2)   {\particle
			{$\bm{1}^+$}        {killed}       {}{}{}{}};
		\node at(3,-3)   {\particle
			{$\bm{2}^+$}    {killed}         {}{}{}{}};
		\node at(4, 0)   {\particle
			{$\bm{2}^+$}        {GW}   {}{}{}{}};
		\node at(4,-1)   {\particle[gray!60!white]
			{$\bm{0}^-$}        {DM?}  {}{}{}{FLRW}};
		\node at(4,-2)   {\particle
			{$\bm{1}^-$}      {killed}        {}{}{}{}};
		\node at(4,-3)   {\particle
			{$\bm{2}^-$}  {killed}    {}{}{}{}};
		
		\draw [mbrace] (-0.8,0.48)  -- (-0.8,-0.48);
		\node at(-1.5,0) [text width =1em]{GR};
		\draw [mbrace] (-0.8,-0.52) -- (-0.8,-3.48);
		\node at(-1.5,-2) [text width =1em]{extra modes in PGG};
		
		\node at(0.5,0.8) {General Case};
		\node at(3.5,0.8) {Minimum};
	\end{tikzpicture}
\caption{``Particle'' spectrum of PGG in general case (left panel) and in our minimum case (right panel). ``?'' indicates that the mode may exist, depending on the Lagrangian parameters. ``Killed'' means the mode has been suppressed. There are three modes remaining on the FLRW background in the minimum case, where two of them are massive.}\label{fig:spectrum}
\end{figure}

\textit{Cosmic dynamics.}---
Under the spatially homogeneous and isotropic reduction, spacetime possesses six \emph{global} symmetries: three spatial translations and three rotations.
Besides, symmetries referred to the temporal direction are \emph{local}.
Six global Killing fields, additionally trivial assumption of spatial topology, lead to that the tetrad field residues one degree of freedom (DOF),
\begin{equation}\label{eq:reduce_tetrad}
	e_0{}^{\hat{0}}=1,~~~~e_i{}^{\hat{j}}=a(t)\delta_{ij},
\end{equation}
i.e. the scale factor $a(t)$, where $t$ is the cosmic time corresponding to the $0$th component, and $i,j,k=1,2,3$ are spatial indices. The hatted indices are decomposed from the Latin alphabet.
Meanwhile, the L-connection residues two DOFs\cite{boehmer2006homogeneous,brechet2008classical}
\begin{equation}\label{eq:phihphif}
	A_i{}^{\hat{j}\hat{0}}=a(t)\phi_h(t)\delta_{ij},~~~~A_i{}^{\hat{j}\hat{k}}=-a(t)\phi_f(t)\epsilon_{ijk},
\end{equation}
i.e. a scalar field $\phi_h(t)$ and a pseudo-scalar field $\phi_f(t)$, corresponding to the spin-$0^+$ and the spin-$0^-$ modes in FIG. \ref{fig:spectrum}, respectively.
Ansatz (\ref{eq:reduce_tetrad}) and (\ref{eq:phihphif}) define the so called Friedmann-Lema\^itre-Robertson-Walker (FLRW) background.
On the FLRW background, the gauge action consisting of the minimum Lagrangian (\ref{eq:Lagrangian}) reduces to
\begin{alignat}{2}
	s^0_G=&\int dtdx^3a^3L^0_G\label{eq:minimum_action}\\
	L^0_{G}=&-6b_0(\dot{\phi}_h+H\phi_h-\phi_h^2+\phi_f^2)-6A_1(\phi_h+H)^2-6A_2\phi_f^2\nonumber\\
	&+B_1\big[(\dot{\phi}_h+H\phi_h)^2-\tfrac{4}{3}(\dot{\phi}_h+H\phi_h)(\phi_h^2-\phi_f^2)\nonumber\\
	&-\tfrac{4}{3}(\dot{\phi}_f+H\phi_f)\phi_h\phi_f+(\phi_h^2-\phi_f^2)^2\big]\nonumber\\
	&+B_2(\dot{\phi}_f+H\phi_f-2\phi_h\phi_f)^2,\label{eq:Lagrangian_0}
\end{alignat}
where dot denotes the derivative with respect to $t$, and $H\equiv\dot{a}/a$ is the Hubble rate.
It can be read from (\ref{eq:Lagrangian_0}) that the Proca masses for $\phi_h$ and $\phi_f$ occur in the forms $\tfrac{1}{2}m_h^2\phi_h^2$, $\tfrac{1}{2}m_f^2\phi_f^2$ with
\begin{equation}\label{def:masses}
	m_h=2\sqrt{3(A_1-b_0)},~~~~m_f=2\sqrt{3(A_2+b_0)}
\end{equation}
as their masses, respectively.

The cosmic dynamic equations corresponding to Lagrangian (\ref{eq:Lagrangian}) take the form of the $\phi$-sourced and $A_1$-rescaled Friedmann equations
\begin{alignat}{4}
H^2=&\tfrac{1}{3A_1}\rho_{\phi},\label{eq0_friedmann}\\
	2\dot{H}+3H^2=&-\tfrac{1}{A_1}p_{\phi},
\end{alignat}
and two equations of motion (Klein-Gordon equations) of $\phi_h$ and $\phi_f$
\begin{widetext}
\begin{alignat}{2}
	\ddot{\phi_h}&+3H\dot{\phi_h}-\tfrac{m_h^2}{4A_1}\phi_h(\dot{\phi_h}+H\phi_h)+2\tfrac{B_1+B_2}{B_1}\phi_f(\dot{\phi_f}+H\phi_f)+\tfrac{1}{2A_1}\phi_h(\tfrac{m_h^2}{2}\phi_h^2+\tfrac{m_f^2}{2}\phi_f^2)+\tfrac{1}{2B_1}\tfrac{\partial V_\phi}{\partial\phi_h}+\tfrac{m_h^2}{2B_1}H=0,\label{eq0_phi_h}\\
	\ddot{\phi_f}&+3H\dot{\phi_f}-\tfrac{m_h^2}{4A_1}\phi_f(\dot{\phi_h}+H\phi_h)-2\tfrac{B_1+B_2}{B_2}\phi_f(\dot{\phi_h}+H\phi_h)+\tfrac{1}{2A_1}\phi_f(\tfrac{m_h^2}{2}\phi_h^2+\tfrac{m_f^2}{2}\phi_f^2)+\tfrac{1}{2B_2}\tfrac{\partial V_\phi}{\partial\phi_f}=0.\label{eq0_phi_f}
\end{alignat}
\end{widetext}
The energy density, the pressure and the potential of $\phi$s are given by
	\begin{alignat}{4}
	\rho_{\phi}=&\tfrac{B_1}{2}(\dot{\phi_h}+H\phi_h)^2+\tfrac{B_2}{2}(\dot{\phi_f}+H\phi_f)^2+\tfrac{1}{2}V_\phi,\label{eq0_krho}\\
	p_{\phi}=&\tfrac{1}{3}[\rho_\phi+\tfrac{m_h^2}{2}(\dot{\phi_h}+H\phi_h-\phi_h^2)-\tfrac{m_f^2}{2}\phi_f^2],\label{eq0_kp}\\
	V_\phi=&\tfrac{m_h^2}{2}\phi_h^2+\tfrac{m_f^2}{2}\phi_f^2-B_1(\phi_h^2-\phi_f^2)^2-4B_2\phi_h^2\phi_f^2.\label{eq0_potential}
\end{alignat}
In addation, it can be checked that the energy density (\ref{eq0_krho}) and the pressure (\ref{eq0_kp}) satisfy the following conservation law
\begin{equation}
	\dot{\rho_{\phi}}=-3H(\rho_{\phi}+p_{\phi}).\label{eq0_rho_phi_point}
\end{equation}
We do the calculations with the help of \texttt{xAct}\footnote{\texttt{xAct: Efficient tensor computer algebra for the Wolfram Language}. Authors: Jos\'e M. Mart\'in-Garc\'ia et. al. Homepage: \url{http://www.xact.es/}}
and integrate our calculations in a Wolfram package \texttt{PGC}\footnote{\texttt{PGC: Symbolic computing package for Poincare Gauge Cosmology}. \texttt{PGC version 1.2.1}: \url{https://github.com/zhanghc0537/Poincare-Gauge-Cosmology}. \texttt{PGC121} is version 1.2.1 of \texttt{PGC}. One can find the field equations in file \texttt{PGC121\_test\_0.nb}, and the FLRW cosmological equations in file \texttt{PGC121\_FLRW-field-equations.nb}, and numerical analysis in file \texttt{PGC121\_numerical-inflation.nb}. \texttt{PGC122} is another version includes the spin projection operator formalism and the calculation of the saturated propagator. Please feel free to download and install our package if you want to check the calculations. The \texttt{README} file will indicate you how to use it.}, which is available on Github.

It is worth mentioning that although we did not add additional material terms to the Lagrangian (\ref{eq:Lagrangian}), the material composition constructed by $\phi_h$ and $\phi_f$ appears on the right side of the Friedmann equations.
From the potential (\ref{eq0_potential}) and equations of motion (\ref{eq0_phi_h},\ref{eq0_phi_f}), it can be seen that there is interaction and momentum exchange between the two fields, with the intensity related to the values of $B_1$, $B_2$. Both fields are up to quartic-order in the potential. Due to $B_1$ being positive, the quartic coefficients are negative. It means that the potential has an inverted Mexican-hat shape, which causes the system to suffer from vacuum instability problem. Fortunately, the $4$th terms in (\ref{eq0_phi_h}) and (\ref{eq0_phi_f}) contribute $\phi_h^3$ and $\phi_f^3$ terms (i.e. external forces) to the KG equations, respectively, and correspond to two additional quartic terms to the potential. Therefore, if the coefficients of the additional quartic terms are greater than the original absolute values, then the system would be stable, which leads to the following constraint conditions on parameters
\begin{equation}
	m_h^2>8A_1, ~~~~m_f^2>8A_1\tfrac{B_1}{B_2}.
\end{equation}

\begin{figure}[htbp]
	\centering
	\subfloat[]
	{\includegraphics[width=0.5\columnwidth]{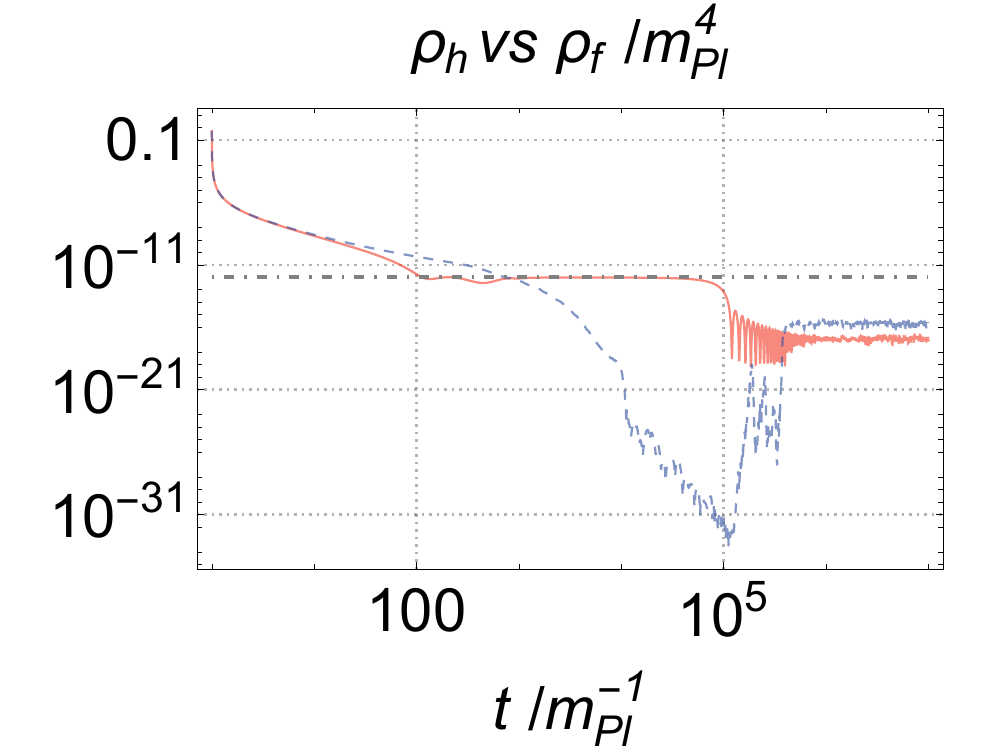}\label{fig:Vhf}}
	\subfloat[]
	{\includegraphics[width=0.5\columnwidth]{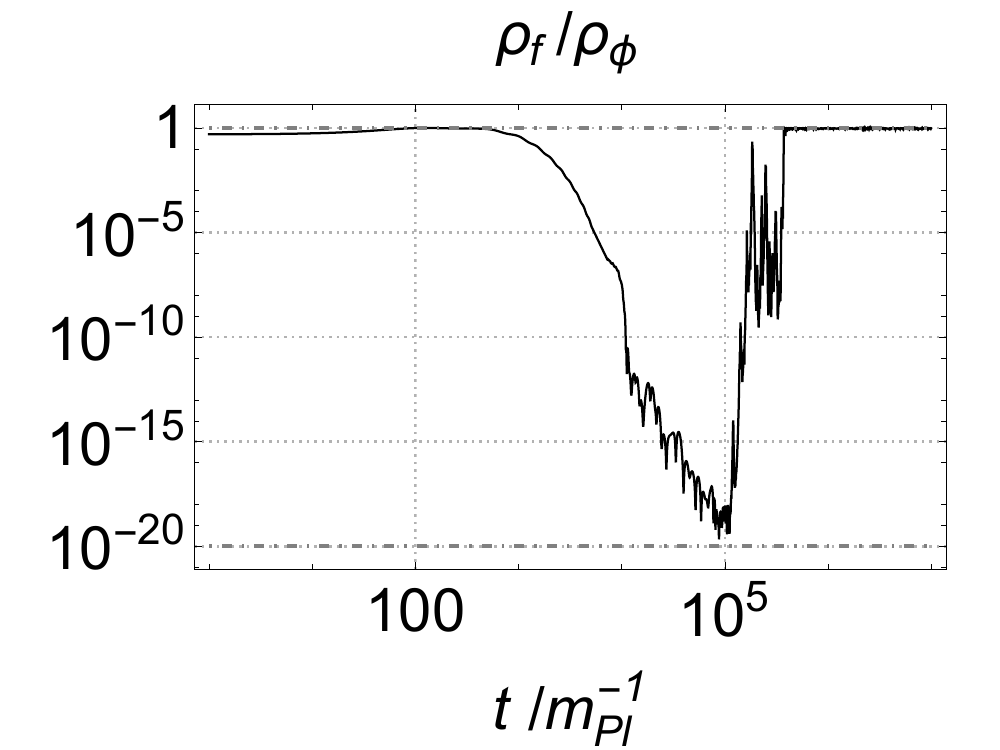}\label{fig:rhohf}}\\
	\subfloat[]
	{\includegraphics[width=0.5\columnwidth]{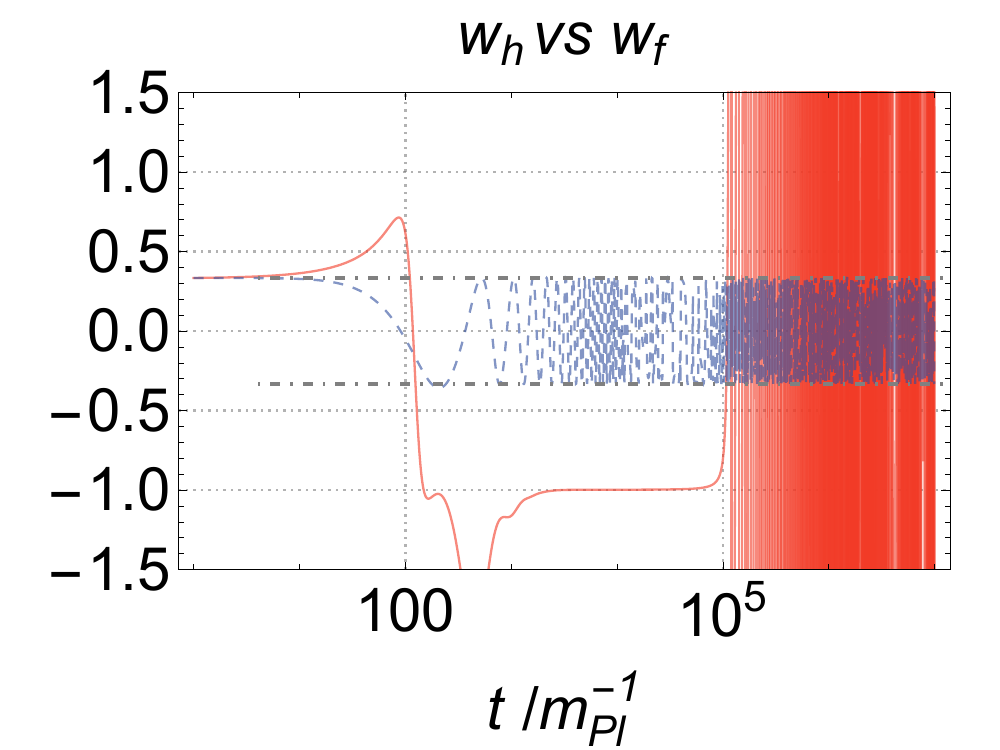}\label{fig:wh}}
	\subfloat[]
	{\includegraphics[width=0.5\columnwidth]{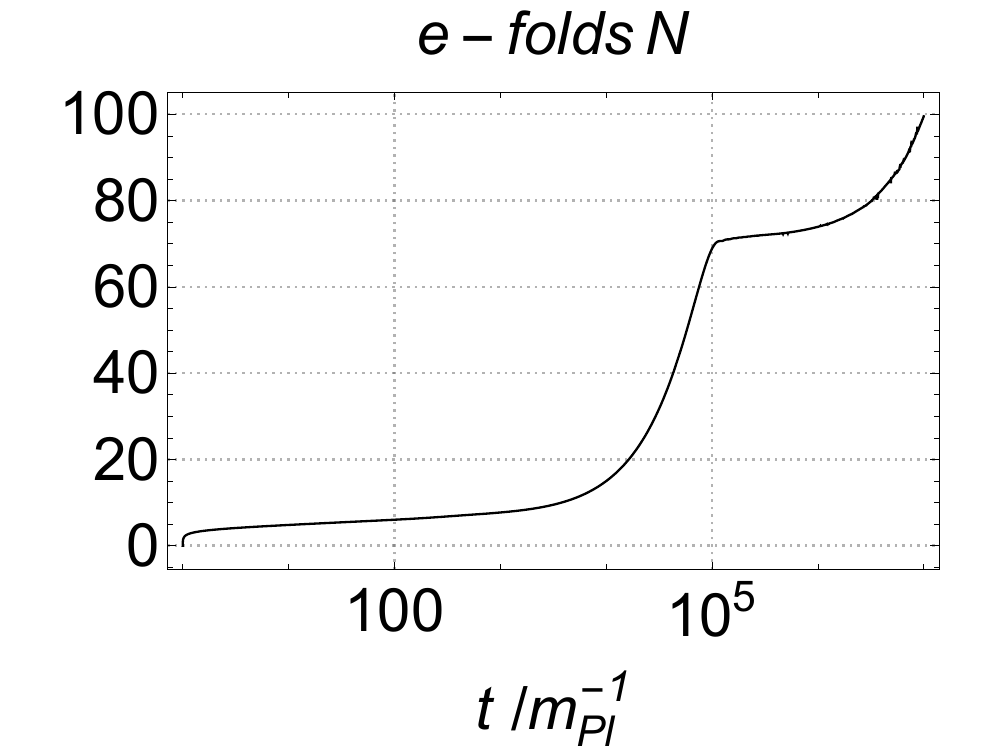}\label{fig:wf}}
	\caption{(a) Evolution of $\rho_h$ (red solid) and $\rho_f$ (blue dash). The gray dot dash line marks $10^{-12}$. (b) Ratio of $\rho_f$ to $\rho_\phi$ (black solid). The upper gray dot dash line marks $1$ and the lower marks $10^{-20}$. (c) Evolution of $w_h$ (red solid) and $w_f$ (blue dash). The two gray dot dash lines mark $\pm1/3$. (d) Evolution of e-folds $N$.}
	\label{fig:total}
\end{figure}

\textit{Inflaton and dark matter candidate.}---
To understand the system intuitively, we tend to do numerical analysis by choosing appropriate values of parameters and initial conditions.
The modern Hot Big Bang theory and the power spectrum of Cosmic Microwave Background (CMB) radiation observation have provided some requirements on models of the very early universe:
\begin{itemize}
\item[1)] the energy density of inflaton should be on $-12$ orders of magnitude during inflation in terms of Planck mass ($m_{Pl}^4$),
\item[2)] the energy density ratio of dark matter is diluted to the order of $-20$ at the end of inflation and approaches $1$ after reheating,
\item[3)] the e-folds during inflation are about $60$, and the reheating process goes through about $20$ e-folds.
\end{itemize}
There are totally 5 parameters in our system. We fix $B_1$ and $B_2$ because they can be rescaled from the action (\ref{eq:minimum_action}). We believe that the initial kinetic energy of the universe is on the Planck scale, so it's reasonable to set the initial values of the fields to $0$ and the initial velocities to $1$, in terms of Planck mass, see (\ref{eq0_krho}). Now we only have three mass-related parameter values to choose in (\ref{def:masses}). To meet the previous requirements, we choose the following parameter values
\begin{alignat}{2}
	A_1=&5.0\times 10^{-7}m_{Pl}^2,~~~~b_0=1.5\times 10^{-8} m_{Pl}^2,\nonumber\\
	A_2=&5.0\times 10^{-6}m_{Pl}^2,~~~~B_1=B_2=1,
\end{alignat}
and initial conditions at Planck time $t_0=1m_{Pl}^{-1}$
\begin{equation}
	\phi_h(t_0)=\phi_f(t_0)=0m_{Pl},~~~~\dot{\phi_h}(t_0)=\dot{\phi_f}(t_0)=1m_{Pl}^2,
\end{equation}
then, the corresponding visualizations are shown in FIG. \ref{fig:total}.
In fact, these three parameter values can fluctuate within certain ranges, and the above choice is a set that we believe meets the requirements quite well. The accurate constraints should be obtained by comparing the primordial power spectrum (PPS) with the actual observation from CMB radiation, which is our subsequent work.

According to the numerical analysis, the interaction between $\phi_h$ and $\phi_f$ is significantly small, so we can ignore the interaction terms in potential (\ref{eq0_potential}) and separate the total energy density $\rho_{\phi}$ (\ref{eq0_krho}) into two parts: $\rho_{h}$ and $\rho_{f}$, and plot them in FIG. \ref{fig:Vhf}.
It is obvious that the whole evolution process can be divided into four periods:
\begin{itemize}
	\item[i)]pre-inflation, $1m_{Pl}^{-1}\sim100m_{Pl}^{-1}$,
	\item[ii)]slow-rolling inflation, $\sim100m_{Pl}^{-1}\sim10^5m_{Pl}^{-1}$,
	\item[iii)]reheating, $\sim10^5m_{Pl}^{-1}\sim10^6m_{Pl}^{-1}$,
	\item[iv)]equilibrium, $>10^6m_{Pl}^{-1}$.
\end{itemize}
Where $\phi_h$ dominates the slow-rolling inflationary period, and spontaneously decays to the reheating period.
During inflation, $\rho_{h}$ is about $-12$ orders of magnitude, which meets the requirement 1). Meanwhile, $\rho_{f}$ is sharply diluted by more than $20$ orders of magnitude, then due to the interaction between $\phi_f$ and $\phi_h$ in the KG equations, $\phi_h$ decays partly to $\phi_f$, causing $\rho_f$ to rebound until it exceeds $\rho_h$ in the equilibrium period.
The ratio of $\rho_f$ to $\rho_\phi$ in FIG. \ref{fig:rhohf} shows clearly that the requirement 2) is also met.
The equation of state defined as $w\equiv p/\rho$ is an important quantity that characterizes the properties of cosmic components. $w<-1/3$ is a necessary condition for causing an accelerating expansion of the universe and $w=0$ marks pressureless non-relativistic matter. We plot $w_h$ and $w_f$ in FIG. \ref{fig:wh}. It shows that $w_h$ is approximately $-1$ during inflation, indicating that $\phi_h$ is indeed a inflaton.
$w_f$ oscillates rapidly between $\pm1/3$ with a period significantly shorter than the dynamical scale we concerned, i.e. its average value is abou $0$.
So far, although we are not yet clear about the specific properties of dark matter particles, to become a candidate, some general limitations need to be met. For example, they must be stable enough on the cosmic time scale so that they can still exist today. In addition, they cannot have strong or electromagnetic interaction.
It has been known that an alternative cold dark matter candidate is a coherently oscillating scalar field, the archetypal example being axion dark matter.
Such coherent scalar fields are therefore a well developed alternative to the weakly-interacting massive particle paradigm\cite{matos2001further,liddle2006inflation}.
$w_f$ shows that $\phi_f$ behaves like pressureless axion matter, and is indistinguishable from traditional cold dark matter candidates on the background level.
We plot e-folds $N\equiv\int Hdt$ in FIG. \ref{fig:wf} which meets the requirement 3).

\begin{figure}[htbp]
	\centering
	\includegraphics[width=\columnwidth]{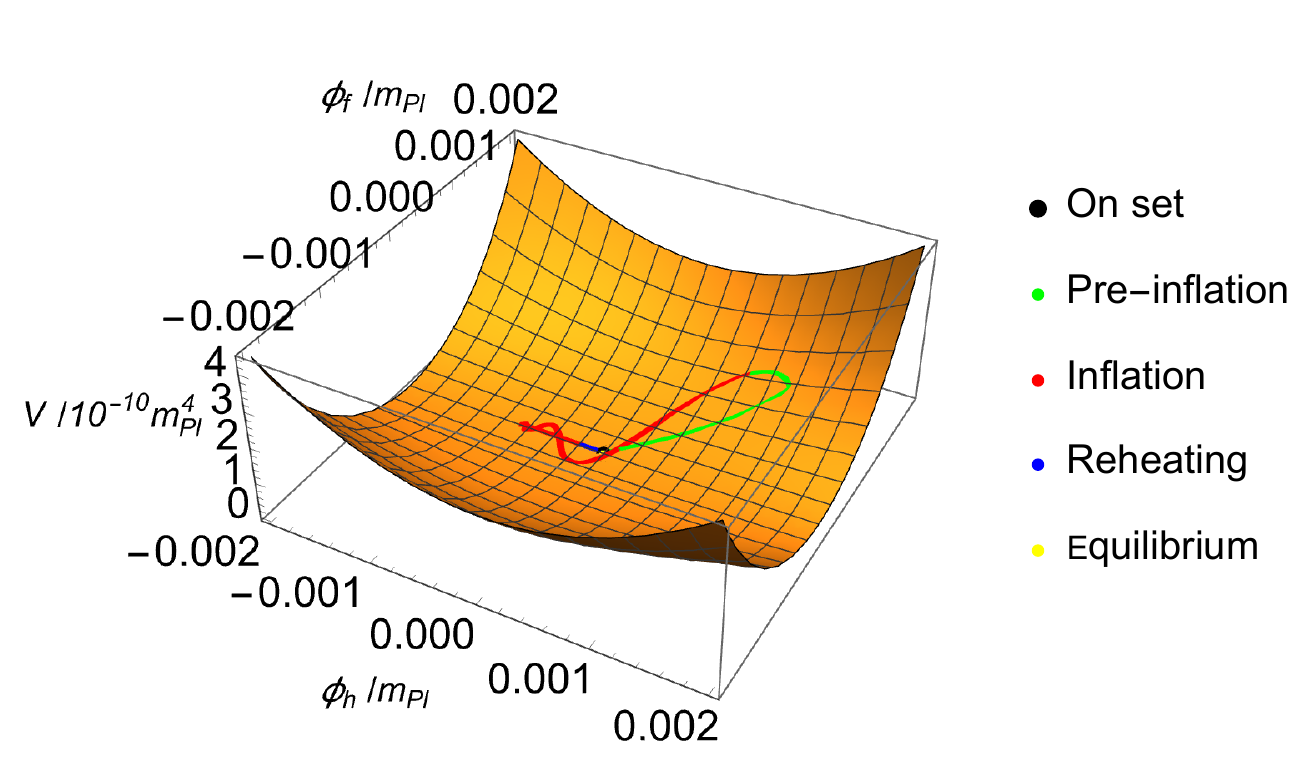}
	\caption{Evolution phase diagram of $\phi_h$ and $\phi_f$ in the effective potential. The evolution trajectory is divided into four stages.}
	\label{fig:V3D1}
\end{figure}

At the end of the Section \textit{Cosmic dynamics}, we have pointed out the relationship between the shape of potential and the stability of system, as well as the restrictions on parameters.
To visually show the stability, we draw the evolution phase diagram of the fields in the effective potential, namely, the potential compensated by the additional terms from KG equations, see FIG. \ref{fig:V3D1}. From the shape of the effective potential, the system is stable.

\textit{Conclusion and discussion.}---
In this letter, we introduce a cosmological model in the framework of PGG, in which cosmological constant, inflaton and dark matter candidate all naturally originate.
Firstly, according to previous studies, in the process of breaking of the P-symmetries down to the L-symmetries, cosmological constant originates in the Lagrangian of matter.
Then we select a ghost- and tachyon-free parity-conserving EH-YM type gauge Lagrangian, which is a minimum extension of GR with two additional massive modes.
By considering the cosmological reduction, the tetrad field residues the scale factor $a$, and the L-connection residues a scalar field $\phi_h$ and a pseudo-scalar field $\phi_f$.
The cosmic dynamic is given by the $\phi$s-sourced and $A_1$-rescaled Friedmann equations.
Numerical analysis shows that $\phi_h$ dominates a slow-rolling inflation and $\phi_f$ behaves as a dark matter candidate.

From the perspective of GR cosmology, $\phi_h$ and $\phi_f$ defined in (\ref{eq:phihphif}) are related to the vector and axial vector components of torsion tensor, respectively
\begin{equation}
	T_{i0}{}^j=(\phi_h+H)\delta_{ij},~~~~T_{ij}{}^k=-2a\phi_f\epsilon_{ijk}.
\end{equation}
These propagating components of torsion can be regarded as the geometric ``substances'' on the background.
``Torsion cannot propagate'' is a misunderstanding brought to us by Einstein-Cartan theory, which is the minimum extension of GR in Riemann-Cartan spacetime with respect to the EH action\cite{obukhov1987weyssenhoff}.
Based on our previous analysis, it can be seen that the missing torsion in most modern theories of gravity plays an important role in the very early universe and the formation of large-scale structures.
From the perspective of gauge theory, $\phi_h$ and $\phi_f$ are just ``gauge bosons of gravity''.
They are similar to the two polarizations of photons, the difference being that photons have no extra DOFs in the direction of propagation, so photons are massless\cite{higgs1964broken}, while in the direction of cosmic evolution, scale factor $a$ plays the role of a Goldstone DOF, which ``gives'' the two gravity bosons masses.

In the subsequent work, we will continue: 1) To obtain the primordial power spectrum of the model through perturbation, so that to constrain the parameters by comparing with observation of CMB radiation; 3) To consider the interactions with standard model particles.
More general cosmology based on gauge theories of gravity beyond the P-group will also be studied in the future.


This work is supported by the China Postdoctoral Science Foundation Grant No. ZD22251090009, the National Natural Science Foundation of China Grant No.12005110, the Natural Science Foundation of Shandong Province, China Grant No. ZR2020QA078,
the Zhejiang Provincial Natural Science Foundation of China Grant No. LR21A050001 and LY20A050002, the National Key Research and Development Program of China Grant No.2020YFC2201503, the National Natural Science Foundation of China Grant No. 11675143 and No. 11975203, and the Fundamental Research Funds for the Provincial Universities of Zhejiang in China Grant No. RF-A2019015.

\bibliography{PGG_SSB_bib}

\end{document}